\documentstyle[12pt,aaspp4]{article} 

\def\gsim{ \lower .75ex \hbox{$\sim$} \llap{\raise .27ex \hbox{$>$}} } 
\def\lsim{ \lower .75ex\hbox{$\sim$} \llap{\raise .27ex \hbox{$<$}} } 

\lefthead{Ghisellini and Celotti}
\righthead{Quasi-thermal Comptonization in gamma-ray bursts}

\begin{document} 

\title{Quasi--thermal Comptonization and gamma--ray bursts}

\author{Gabriele Ghisellini} 
\affil{Osservatorio Astronomico di Brera, Via Bianchi 46, 
I--23807 Merate, Italy}
\authoremail{gabriele@merate.mi.astro.it} 

\and 

\author{Annalisa Celotti}
\affil{S.I.S.S.A., Via Beirut 2/4, I--34014 Trieste, Italy}
\authoremail{celotti@sissa.it}

\begin{abstract}
Quasi--thermal Comptonization in internal shocks formed between
relativistic shells can account for the high energy emission of
gamma--ray bursts.  This is in fact the dominant cooling mechanism if
the typical energy of the emitting particles is achieved either
through the balance between heating and cooling or as a result of
electron--positron pair production. Both processes yield sub or mildly
relativistic energies.  In this case the synchrotron spectrum is
self--absorbed, providing the seed soft photons for the Comptonization
process, whose spectrum is flat [$F(\nu)\sim$const], ending either in
an exponential cutoff or a Wien peak, depending on the scattering
optical depth of the emitting particles. Self-consistent particle
energy and optical depth are estimated and found in agreement with the
observed spectra.

\end{abstract}

\keywords{gamma rays: bursts --- radiation mechanisms: non--thermal,
thermal --- X-rays: general}

\section{Introduction}

After the observational breakthrough of $Beppo$SAX (Costa et al. 1997;
Van Paradijs et al. 1997) we are now starting to disclose the physics
of gamma--ray bursts (GRB).  The huge energy and power release required by
their cosmological distances supports the fireball scenario (Cavallo
\& Rees 1978; Rees \& M\'esz\'aros 1992; M\'esz\'aros \& Rees 1993),
whose evolution and behavior is unfortunately largely independent of
its origin.

We do not know yet in detail how the GRB event is related
to the afterglow emission, but in the most accepted scheme of
formation of and emission from internal/external shocks (Rees \&
M\'esz\'aros 1992; Rees \& M\'esz\'aros 1994; Sari \& Piran 1997), the
former is due to collisions of pairs of relativistic shells (internal
shocks), while the latter is generated by the collisionless shocks
produced by shells interacting with the interstellar medium (external
shocks). The short spikes ($t_{\it var}\sim$10 ms) observed in the
high energy light curves suggest that shell--shell collisions occur at
distances $R\simeq 10^{12}$--$10^{13}\ $ cm from the central source
within a plasma moving with a bulk Lorenz factor $\Gamma \ge 100$. The
fireball starts to be decelerated by the interstellar medium further
out, at a distance which depends on the assumed density of this
material.

Up to now, the main radiation mechanism assumed to give rise to both
the burst event and the afterglow is synchrotron emission (Rees \&
M\'esz\'aros 1994; Sari, Narayan \& Piran 1996; Sari \& Piran 1997;
Panaitescu \& M\'esz\'aros 1998).  In fact if magnetic field,
electrons and protons share the available energy $E=10^{50} E_{50}$
erg of the shell, the electrons reach typical random Lorentz factors
$\gamma\sim m_p/m_e$, while the assumption of a Poynting flux
$L_B=R^2\Gamma^2B^2c/2 = 10^{50}L_{B,50}$ erg s$^{-1}$ implies a
comoving magnetic field of the order $B\sim L_B^{1/2}/(R\Gamma)\sim
10^5 L_{B,50}^{1/2}/(R_{13}\Gamma_2)$ G. \footnote{Here and in the
following we parametrise a quantity $Q$ as $Q=10^xQ_x$ and adopt cgs
units.} 
For these values of $\gamma$ and $B$, the typical observed
synchrotron frequency is $\nu_s \sim 0.5 L_{B,50}^{1/2}/[R_{13}(1+z)]$
MeV, independently of the bulk Lorentz factor $\Gamma$, and in
excellent agreement with the observed values of the peak of GRB
spectra in a $\nu F(\nu)$ representation.

In this model therefore the assumption of energy equipartition plays a
key role.  And, at a closer look, this implicitly requires a number of
constraints to be satisfied in the emission regions.  The main one
concerns the acceleration of the electrons, which must be impulsive
(i.e. on timescales much shorter than the cooling ones). Further
requirements will be discussed in \S 2.

One can envisage an alternative scenario, which we describe in \S 3,
where the key role is instead played by the balance between the
cooling and heating processes. This would be favoured if the emitting
region occupies the entire shell volume rather than the narrow region
associated with a planar shock, as in the `equipartition' model.

An immediate prediction following this hypothesis is that the typical
energy of the emitting electrons is mildly relativistic, and the main
radiation process is quasi--thermal Comptonization.  Also this model
implies the existence of a characteristic observed frequency of a few
MeV, controlled by the feedback introduced by the effect of
electron--positron pair production.

Earlier attempts to explain burst radiation with multiple Compton
scatterings have been done by Liang et al. (1997), who considered
emission by a very dense population of non--thermal relativistic
electrons in a weak magnetic field, in the context of bursts located
in an extended galactic halo.  Liang (1997) later extended this model
for bursts at Gpc distances, assuming an emitting region of the order
of $\sim 10^{15}$ cm, a magnetic field value of the order of 0.1 G and
thermal electrons at a temperature of a few keV.

We instead first discuss the constraints that the `synchrotron'
scenario must satisfy and then, using {\it the very same parameters}
(except for the electron energy), propose that
quasi--thermal Comptonization can quite naturally dominate the
cooling.

Some consequences of our scenario are presented in \S 4, while
our findings are discussed in \S 5.

For simplicity we will consider spherical shells of comoving width
\footnote{Primed quantities are measured in the comoving frame.  The
magnetic field, the random Lorentz factor $\gamma$, the optical depth
$\tau$ and the Comptonization parameter $y$, not primed for clarity,
are also measured in the comoving frame.}  $\Delta R^\prime$, moving
with a bulk Lorentz factor $\Gamma$, and a monoenergetic electron
distribution peaked at a (random) energy $\gamma m_e c^2$.
Generalizations to jet--like outflows and more complex particle
distributions are straightforward.

\section{Constraints on the `equipartition' scenario}

Shell--shell collision produces a shock that has to accelerate
particles to an energy $\gamma m_{\rm e} c^2$ in a timescales
$t^\prime_{\rm acc}$ much shorter than the cooling time $t^\prime_{\rm
cool}(\gamma)$, since otherwise the final random energy of the
particle would be controlled by its cooling rate and not by the
equipartition condition.

Once accelerated to the equipartition Lorentz factor $\gamma\sim
m_p/m_e$ and radiatively cooled down, a particle will not be
accelerated again during the shell--shell interaction, in order for
the total random energy not to exceed the available energy in the
relative bulk motion of the two shells.

The inverse Compton power must be at most of the same order of the
synchrotron one, as otherwise the luminosity we observe in the hard
X--rays would be largely dominated by the luminosity emitted beyond
the GeV band, with obvious problems for the total energetics.  This
requires that $\tau_e\gamma^2\le 1$, where $\tau_e$ is the optical
depth of the emitting relativistic electrons (not to be confused with
the total scattering optical depth $\tau_T$ of the shell which is of
order unity at $R\sim 10^{13}$ cm).  This condition is valid if the
Compton scattering process is entirely in the Thomson regime, which is
appropriate as long as $\gamma\le 760 B_5^{-1/3}$.  Higher energy
electrons would scatter in the Thomson limit synchrotron photons of
lower frequencies: for $\gamma\sim 10^3$ the reduction in the Compton
luminosity due to Klein--Nishina effects is of the order of $3\,
B_5^{4/3}\gamma_3^4$.

The condition $\tau_e\gamma^2\le 1$ translates in demanding that the
width $dR^\prime$ of the emitting region of the shock (assumed to be
plane--parallel) satisfies $dR^\prime/\Delta R^\prime\le
\tau_T^{-1}\gamma^{-2} \sim 10^{-6}$.  This in turn controls the
cooling timescale, which must be of the order of $dR^\prime/c$, and we
obtain:
\begin{equation}
t^\prime_{cool} \, \le \, {\Delta R^\prime \over c \,\tau_T \gamma^2}
\, \to \, \gamma\, \le\, {\Delta R^\prime \over \tau_T}\, {\sigma_T
B^2 (1+\tau_e\gamma^2) \over 6\pi m_ec^2}\, \sim 86 \, { \Delta
R^\prime_{11} B_5^2 \over \tau_T}
\end{equation}
indicating that either the magnetic field is stronger than $10^5$ G or
the total shell optical depth $\tau_T\sim 0.1$, to allow the electrons
to reach $\gamma\sim 10^3$.

Consider also that if the density of electrons is increased by
electron--positron pair production, the mean energy per lepton is less
than $m_p/m_e$ by the factor equal to the ratio of leptons to proton
densities $n^\prime_e/n^\prime_p$.

We consider these constraints -- i.e. impulsive acceleration, limited
Compton power and absence of copious pair production -- quite
demanding (at least among the limits imposed by considerations on
radiation processes).  The emitting region cannot be very compact and
cannot have a width larger than $dR^\prime$.  Note that in some models
the shocked region is instead considered to be complex and extended,
as in the shock structure resulting from Rayleigh-Taylor instability,
which can occupy the entire shell volume (Pilla \& Loeb 1998).  In
this case there are many more electrons emitting at a given time, with
a consequent building up of the radiation energy density and an
increased Compton luminosity.  In addition, the larger cooling rate
can limit the typical electron energies to values much below the
equipartition one.

In the next section we will therefore examine the possibility that the
above conditions are not satisfied.

\section{Quasi--thermal Comptonization}

Let us assume that the heating process for a typical electron lasts
for the duration time of the shell--shell interaction, $\sim \Delta
R^\prime/c$.  The maximum amount of energy given to a single lepton is
of the order of $m_p c^2 n^\prime_p/n^\prime_e$ (not to violate the
total energetics), which, when released over the above timescale
corresponds to a total (average) heating rate $\dot E^\prime_{heat} =
n^\prime_p m_p c^3/\Delta R^\prime$.  The typical electron energy is
given by balancing $\dot E^\prime_{heat}$ and $\dot E^\prime_{cool}=
(4/3) n^\prime_e\sigma_T c U^\prime_r \gamma^2\beta^2
(1+U^\prime_B/U^\prime_r)$, where $U^\prime_B/U^\prime_r$ is the
magnetic to radiation energy density ratio and where $n^\prime_e$
includes a possible contribution from $e^\pm$ pairs:
\begin{equation}
\gamma^2\beta^2 \, \approx \, { 3\pi R^2 n^\prime_p m_pc^3 \over
\Delta R^\prime \sigma_Tn^\prime_e L^\prime
(1+U^\prime_B/U^\prime_r)}\, \equiv \, {n^\prime_p m_p\over n^\prime_e
m_e}\,{1\over 1+U^\prime_B/U^\prime_r}\, {3\pi \over \ell^\prime}
\end{equation}
where $\ell^\prime\equiv [L^\prime\sigma_{\rm T}/(Rm_{\rm e}c^3)]
(\Delta R^\prime/R)$ is the compactness parameter of the region
emitting a (comoving) luminosity $L^\prime$.  Electron--positron pairs
can be important for values of the compactness greater than unity
(Svensson 1982, 1984, 1987), and can even dominate the particle number
density.  A typical value in this situation is $\ell^\prime = 270 \,
(L^\prime_{46}/ R_{13})(\Delta R^\prime_{11} /R_{13})$, and therefore
equation (2) yields typical electron (and positron) energies at most
moderately relativistic.
As detailed below, the small energy of the emitting particles implies:

1) the synchrotron emission is self--absorbed.

2) the main radiation mechanism is multiple Compton scattering and the
self--absorbed synchrotron emission is the source of soft seed
photons.

Even though the particle distribution may not have time to thermalize,
it will be characterized by a mean energy, and possibly be peaked at
this value.  It is then convenient to introduce an 'effective
temperature' $\Theta^\prime\equiv kT^\prime/(m_ec^2)$.

The synchrotron luminosity can then be estimated assuming that the
spectrum is described by the Rayleigh-Jeans part of a blackbody
spectrum, up to the self--absorption frequency $\nu_T^\prime$:
\begin{equation}
L^\prime_s \, \sim {8\pi \over 3} \, m_e R^2\Theta^\prime
(\nu_T^\prime)^3 \, \sim 7.6\times 10^{41} \Theta^\prime R_{13}^2
(\nu_{T,14}^\prime)^3 \,\, {\rm erg~s^{-1}}
\end{equation}
where $\nu_T^\prime$ can be derived again approximating the particle
distribution as a Maxwellian of temperature $\Theta^\prime$. An
approximate prescription, which holds for $0.1 \lsim \Theta^\prime
\lsim 3$, has been derived by interpolating the analytic
approximations to the cyclo--synchrotron emission reported by
Mahadevan et al. (1996). This gives, for $B_5\sim 1$ and $\tau_T\sim
1$, $ \nu_{T}^\prime \, \sim 2.75\times 10^{14}
(\Theta^\prime)^{1.191}\,\, {\rm Hz}.  $

A generalized Comptonization parameter $y$, which is approximately
valid also in the trans--relativistic and quasi--transparent
conditions, can be defined as $ y\, \equiv\, 4\tau_T\Theta^\prime
(1+\tau_T)(1+4\Theta^\prime).  $

\noindent
The ratio of the Compton to the synchrotron powers can then be
approximated by $e^y$, and thus in order to emit a Compton comoving
luminosity $L^\prime_c = 10^{46}L^\prime_{c,46}$ erg s$^{-1}$, the $y$
parameter must be of the order of $y= \ln(L^\prime_c/L^\prime_s)\sim
11.5 \,\ln(L^\prime_{c,46}/L^\prime_{s,41})$.  With this value of $y$
and $\tau_T$ of order unity, the Comptonized high energy spectrum has
a $F(\nu)\propto \nu^0$ shape, while the relatively modest optical
depth prevents a strong Wien peak to form (Pozdnyakov, Sobol \&
Sunyaev 1983).  Therefore, very schematically, the resulting observed
spectrum would extend between the energies $h\nu_T^\prime\Gamma/(1+z)$
and $\sim 2 kT^\prime\Gamma/(1+z)$, with $F(\nu)\sim $ const.

The spectrum emitted by a single shell will evolve very rapidly: after
the observed acceleration time $\Delta R^\prime (1+z)/(\Gamma c)$,
particles cool on a similar timescale, while the Comptonization
spectrum steepens and the emitted power decreases.  (Eventually, the
same particles can be re--heated by a collision with another shell.)
The time integrated emission (even for a short exposure time of -- say
-- a second) will result from all the cooling and re--heating
histories of many shell--shell interactions.  Any Wien hump and/or
feature in the spectrum of individual shells will be smoothed out.
The hard power--law continuum, if typical of all shells, would instead
be preserved even when integrating over the exposure time.

\subsection{The role of electron--positron pairs}

As anticipated, $e^{\pm}$ pair production can play a crucial role:
this process would surely be efficient for intrinsic compactnesses
$\ell^\prime>1$, and would on one hand increase the optical depth, and
on the other act as a thermostat, by maintaining the temperature in a
narrow range.  For the temperatures of interest here, photon--photon
collisions are the main pair production process.  Note that our
definition of $\ell^\prime$ corresponds to the optical depth for
$\gamma$--$\gamma\to e^\pm$ {\it within the shell width}.  Additional
pairs will be produced outside the shell region, increasing the lepton
content of the surrounding medium.  Detailed time dependent studies of
the optical depth and temperature evolution for a rapidly varying
source have not yet been pursued.  Results concerning a steady source
in pair equilibrium 
indicate that for $\ell^\prime$ between 10 and $10^3$ the maximum
equilibrium temperature is of the order of 30--300 keV (Svensson 1982,
1984), if the source is pair dominated (i.e. the density of pairs
outnumbers the density of protons). Indeed we expect in this situation
to be close to pair equilibrium, as this would be reached in about a
dynamical timescale.  Note that the quoted numbers refer to a perfect
Maxwellian particle distribution. However, pairs can be created even
if the temperature is sub--relativistic by the photons and particles
in the high energy tails of the real distribution: for a particle
density decreasing slower than a Maxwellian one, more photons are
created above the threshold for photon--photon pair production, and
thus pairs become important for values of $\Theta^\prime$ lower than
in the completely thermal case.

We conclude that an `effective' temperature of $kT^\prime\sim$ 50 keV
($\Theta^\prime\sim 0.1$) and $\tau_T\sim 4$ dominated by
pairs, can be a consistent solution giving $y\sim 11$.  We stress here
that the assumption of a soft seed photon distribution of luminosity
$L^\prime\sim 10^{41}$ erg s$^{-1}$ implies that $any$
self--consistent solution $must$ give $y\sim 10$, in order to produce
a Compton luminosity matching the observed one.

\section{Some consequences}

If the high energy spectrum is due to quasi--thermal Comptonization,
it will be sensitive to the amount of available soft seed photons.
Assume that the complex light curve of GRB can be
explained by the internal shock scenario, in which the emission is
produced by the collisions of many shells.  The first colliding shells
will give rise to a certain synchrotron self--absorbed radiation,
while subsequent shells, besides producing synchrotron
radiation, will be illuminated by photons coming from the previous
shell--shell collisions.  The amount of seed photons is then bound to
increase, in turn increasing the cooling rate of the
electrons and positrons, which will therefore reach a lower
temperature and produce a softer spectrum.  This can qualitatively
explain the hard--to--soft behavior of GRB emission and
the (weak) correlation between duration and hardness (shorter bursts
have harder spectra, Fishman \& Meegan, 1995 and references therein).
More quantitative details demand the knowledge of the exact time
dependent feedback introduced by pair emission, which is difficult to
asses.

If the intrinsic effective temperature is of the order of 50 keV, then
the observed Comptonized spectrum extends to $\sim 10
\Theta^\prime_{-1}\Gamma_2/(1+z)$ MeV.  Note that there are no severe
constraints on such high values of the energy at which the $\nu
F(\nu)$ spectrum of the burst peaks (Cohen, Piran \& Narayan 1998),
and these values maybe reached during the very first parts of the
burst light curve (e.g. first second).  Time integrated
spectra may instead be well fitted as emission from particles of lower
temperatures.

One would also expect that (again especially during the initial first
phase) a Wien peak at the electron temperature would be formed for
bursts with $\tau_T$ larger than 3--5. This values of the optical
depth would be easily reached for particularly strong bursts, where
more copious pair production might occur.

In Comptonization models, photons of higher energies undergo more
scatterings: if variability is caused by a change in the seed photon
population, this may cause the high energy flux to lag that at softer
frequencies.  In this case the relevant timescale is the average time
between two scatterings, which in the observer frame is of the order
of $\Delta R^\prime (1+z)/(\Gamma \tau_T c)\sim 0.03 \Delta
R^\prime_{11}(1+z)/(\tau_T \Gamma_2)$ s, which is within the
possibility of current detectors.  However there can be no lag if
variability is caused by a sudden increase (decrease) of the number of
emitting electrons or of their effective temperature.  Constraints on
the Comptonization scenario may come from detailed studies on how to
reproduce very fast variability, since multiple scatterings will tend
to smooth out any very short change.

If the progenitors of GRB are hypernovae (Paczy\'nski
1998) the density in the vicinity of the central source is dominated
by the pre--hypernova wind. This can lead to optical depths around
unity at distances $R\sim 10^{12}$--$10^{13}$, just where shell--shell
collisions are assumed to take place.  There is then the possibility
that the GRB events are due to shocks with this material,
rather than shocks between the shells.  The implications of this
scenario for the emission models are very interesting and will be
discussed elsewhere (Ghisellini et al., in prep.). Here we would like
to stress that:

i) in the case of shocks between shells and the pre--hypernova wind
the large densities involved suggest that inverse Compton emission is
favored with respect to the synchrotron process;

ii) if the (still unshocked) interstellar material has total optical
depth $\tau_T$ around or greater than unity, photons will be
down--scattered, introducing a break in the emergent spectrum at the
observed energy $511/[\tau_T^2\,(1+z)]$ keV (Guilbert, Fabian \& Rees
1983; Pozdnyakov, Sobol \& Sunyaev 1983). Furthermore, the
interstellar matter will act as a `mirror', sending back the scattered
photons, thus increasing the amount of Compton cooling in the emitting
region (see Ghisellini \& Madau 1996 for an application of this
`mirror' model to blazars).

We then conclude that the high density environment of the hypernova
poses problems to the non--thermal `equipartition' scenario and favors
quasi--thermal Comptonization as the main radiation process.

Afterglow emission would start at the deceleration radius $R_d\sim
10^{16} [E_{50}/(n \Gamma^2_2)]^{1/3}$ cm, after $\sim$40
$(E_{50}/n)^{1/3}/\Gamma_2^{8/3}$ s from the start of the burst 
(e.g. Wijers, Rees \& M\`esz\`aros 1997), where $n$ is the number
density of the interstellar material.  At $R_d$, the reduced densities
of particles and photons diminish the importance of the Compton
emission even for ultrarelativistic electron energies, but the
requirement $\tau_e\gamma^2<1$ can still limit $\gamma$,
especially in the case of dense, star forming, environments.

\section{Discussion}

In the equipartition scenario, the rough equality between the energy
density of protons, leptons and magnetic field leads to a remarkable
good agreement with the observed characteristics of the spectra.  In
order for this to be achieved the acceleration of electrons has to be
impulsive, take place in a very limited volume of the interacting
shell, and e$^{\pm}$ pair density has to be small enough not to
significantly lower the mean lepton energy.

At the other extreme, when the particle acceleration occupies the
entire shell volume and/or lasts for a shell light crossing time, the
mean lepton energy is controlled by the balance between the heating
and the cooling rate.  This leads to mildly or sub relativistic lepton
energies.  Therefore the synchrotron emission is inhibited by
self--absorption, and provides soft photons for the dominant inverse
Compton scattering process.  The ratio of the observed multiple
Compton scattering and the self--absorbed synchrotron luminosities is
of the order of $10^5$, and determines the required Comptonization $y$
parameter, i.e. the product of the particle optical depth and
temperature.  As long as the compactness of the emission region is
greater than unity, relativistic temperatures cannot be achieved,
because in this case electron--positrons are copiously produced,
increasing the optical depth and decreasing the temperature.  If, on
the other hand, the temperature is low and the optical depth is large,
photons are trapped inside the shell, and part of the radiation
luminosity is used to expand it.  As a result, there is a narrow range
of optical depths and temperatures which accounts for the observed
spectra.  This may be why GRB preferentially emit at
$\sim$1 MeV.

Detailed analysis are needed to determine the exact shape and the time
behavior of the predicted spectra: particularly relevant, in this
respect, will be to study the time evolution of hot compact sources,
relaxing the assumption of pair balance.

We conclude that there are at least two possible regimes yielding the
observed spectrum and peak frequency of GRB, depending
on the nature of the dissipation/acceleration mechanism.  There is
even the possibility that they coexist: one can in fact imagine that
an initially planar (and 'narrow') shocked region can be soon
subjected to instabilities thus starting dissipating energy over a
larger volume. In this situation both the non--thermal `equipartition'
and the quasi--thermal `heating/cooling balance' regimes would be at
work.

\acknowledgements We thank Martin Rees for invaluable discussions and
comments on the manuscript.  AC acknowledges the Italian MURST for
financial support.

\end{document}